\documentclass[12pt,a4paper]{article}
\usepackage{a4}
\usepackage[dvips]{graphics}
\def\units#1{\hbox{$\,{\rm #1}$}}
\def\degrees{\hbox{${}^\circ$}}

\begin{document}


\hskip 250 pt {\bf }




\vspace{1cm}

      
\vskip 40 pt
\begin{center}
{\Large\bf 
Measurement of the energy spectrum of underground muons at Gran Sasso with a
transition radiation detector} \\

\vspace{1cm}

{ \bf The MACRO Collaboration } \\

\nobreak\bigskip\nobreak
M.~Ambrosio$^{12}$, 
R.~Antolini$^{7}$, 
C.~Aramo$^{7,n}$,
G.~Auriemma$^{14,a}$, 
A.~Baldini$^{13}$, 
G.~C.~Barbarino$^{12}$, 
B.~C.~Barish$^{4}$, 
G.~Battistoni$^{6,b}$, 
R.~Bellotti$^{1}$, 
C.~Bemporad$^{13}$
% ...
\\
\vspace{1.5 cm}
\footnotesize
1. Dipartimento di Fisica dell'Universit\`a di Bari and INFN, 70126 
Bari,  Italy \\
2. Dipartimento di Fisica dell'Universit\`a di Bologna and INFN, 
 40126 Bologna, Italy \\
3. Physics Department, Boston University, Boston, MA 02215, 
USA \\
4. California Institute of Technology, Pasadena, CA 91125, 
USA \\
5. Department of Physics, Drexel University, Philadelphia, 
PA 19104, USA \\
6. Laboratori Nazionali di Frascati dell'INFN, 00044 Frascati (Roma), 
Italy \\
7. Laboratori Nazionali del Gran Sasso dell'INFN, 67010 Assergi 
(L'Aquila),  Italy \\
8. Depts. of Physics and of Astronomy, Indiana University, 
Bloomington, IN 47405, USA \\
9. Dipartimento di Fisica dell'Universit\`a dell'Aquila  and INFN, 
 67100 L'Aquila,  Italy \\
10. Dipartimento di Fisica dell'Universit\`a di Lecce and INFN, 
 73100 Lecce,  Italy \\
11. Department of Physics, University of Michigan, Ann Arbor, 
MI 48109, USA \\	
12. Dipartimento di Fisica dell'Universit\`a di Napoli and INFN, 
 80125 Napoli,  Italy \\	
13. Dipartimento di Fisica dell'Universit\`a di Pisa and INFN, 
56010 Pisa,  Italy \\	
14. Dipartimento di Fisica dell'Universit\`a di Roma ``La Sapienza" and INFN, 
 00185 Roma,   Italy \\ 	
15. Physics Department, Texas A\&M University, College Station, 
TX 77843, USA \\	
16. Dipartimento di Fisica Sperimentale dell'Universit\`a di Torino and INFN,
 10125 Torino,  Italy \\	
17. Also  Faculty of Sciences, University Mohamed I, B.P. 424 Oujda, Morocco \\
$a$ Also Universit\`a della Basilicata, 85100 Potenza,  Italy \\
$b$ Also INFN Milano, 20133 Milano, Italy\\
$c$ Also Istituto TESRE/CNR, 40129 Bologna, Italy \\
$d$ Also Scuola Normale Superiore di Pisa, 56010 Pisa, Italy\\
$e$ Also Dipartimento di Fisica, Universit\`a di Roma Tre, Roma, Italy \\
$f$ Also Universit\`a di Trieste and INFN, 34100 Trieste, 
Italy \\
$g$ Also Dipartimento di Energetica, Universit\`a di Roma, 
 00185 Roma,  Italy \\
$h$ Also Institute for Nuclear Research, Russian Academy
of Science, 117312 Moscow, Russia \\
$i$ Also Institute for Space Sciences, 76900 Bucharest, Romania \\
$l$ Swarthmore College, Swarthmore, PA 19081, USA\\
$m$ RPD, PINSTECH, P.O. Nilore, Islamabad, Pakistan \\
$n$ Also INFN Catania, 95129 Catania, Italy\\
$o$ Also Department of Physics, Pennsylvania State University, 
University Park, PA 16801, USA\\
\end{center}
$*$ Corresponding authors: \\
M.N.Mazziotta, e-mail: mazziotta@ba.infn.it \\
P.Spinelli, e-mail: spinelli@ba.infn.it 

\date{ }

\begin{abstract}
We have measured directly the residual energy of cosmic ray muons crossing 
the MACRO detector at the Gran Sasso Laboratory. For this measurement
we have used a transition radiation detector consisting of three 
identical modules, each of about 12\units{m^2} area, operating in the 
energy region from 100\units{GeV} to 1\units{TeV}.  
The results presented here were
obtained with the first module collecting data for more than two years. 
The average single muon energy is found to be
320 $\pm$ 4 (stat.) $\pm$ 11 (syst.)\units{GeV} in the rock depth 
range 3000--6500\units{hg/cm^2}.
The results are in agreement 
with calculations of the energy loss of muons in 
the rock above the detector.

\end{abstract}


\begin{center}
{To be submitted to Astr. phys.}
\end{center}



\newpage

\section*{Introduction}
High energy muons are produced in interactions of primary cosmic rays 
with nuclei in the
Earth's atmosphere. The muon energy distribution is dependent 
on the spectrum and
composition of the primary cosmic rays, and can be used to 
obtain information concerning
these quantities.  In particular, a direct measurement of the 
single muon spectra obtained
deep underground can, in principle, provide 
information about the ``all nucleon'' cosmic ray 
spectra at high energies. This paper describes a measurement
of the high energy underground muon spectrum, carried out 
using a transition radiation detector
(TRD) in association with the MACRO detector.

An attempt was made in 1987 \cite{Cali} to measure the residual 
energy of muons reaching the Mont Blanc
underground laboratory. In this case, a small transition radiation 
detector (TRD) installed on 
the top of the NUSEX detector \cite{Caste} provided the measurement of 
the muon energy in the range 100--500\units{GeV}. 
The measured spectrum was consistent with a surface 
muon differential distribution of the
type $E^{-3.71}$ folded with absorption in 5000\units{hg/cm^2} standard rock. 
More recently, a 
measurement of the cascade showers produced by underground muons 
inside the NUSEX 
calorimeter ~\cite{Cast1,Cast2} was used to obtain an average muon 
energy of 346 $\pm$ 14 $\pm$ 17\units{GeV} at 
a depth of 5000\units{hg/cm^2}. The residual energy spectrum 
was reported to be ``not in 
contradiction with a power law integral distribution 
with an index $\gamma$=2.7--2.9''. 

To expand on these measurements, we have designed 
and built a large area 
TRD, for use in conjunction with the MACRO detector at the Gran 
Sasso Laboratory. The TRD 
allows the energy measurement of muons up to $\sim$ 1\units{TeV}, 
although with modest 
resolution. With this technique the energy of downgoing and 
of neutrino induced upgoing 
muons is measured directly. This allows the local 
spectrum and the average energy versus 
depth to be evaluated, independent of assumptions 
on the particle zenith angle distribution 
and of the energy losses in the surrounding rock \cite{Gsro}.  

\section{The MACRO TRD}
\subsection{Properties of Transition Radiation}

Transition radiation detectors are presently of interest 
for fast particle identification,
both in accelerator experiments \cite{Dolg} and in cosmic 
ray physics [7-14]. In particular, TRDs
have been proposed and developed to measure the energy of 
cosmic ray muons in the TeV region. The
characteristic dependence of transition radiation on the 
Lorentz factor $\gamma$ of the incident
particle makes it possible to evaluate the energy 
$E=m_0 \gamma c^2$ of the particle if the 
rest mass $m_0$ is known, as it is the case of 
atmospheric muons reaching an underground 
laboratory. TRDs can provide an energy 
measurement of particles over an energy 
range typically spanning one order of magnitude, between 
the transition radiation threshold 
and saturation energy values.

Transition radiation (TR) is emitted in the X-ray region whenever an 
ultrarelativistic charged particle crosses the boundary of 
two materials with different 
dielectric properties \cite{Ginz,Gari}. At each interface the 
emission probability for an 
X-ray photon is of the order of $\alpha=1/137$. Radiators consisting of 
several hundred 
regularly spaced foils are used to enhance X-rays production, allowing 
a reliable tagging of the fast particle.

The ``multilayer'' radiator introduces important physical 
constraints on the radiation 
yield, because of so-called ``interference effects''. It has been 
established that the 
radiation emission threshold occurs at a Lorentz factor 
$\gamma_{th}=2.5 \omega_p d_1$, 
where $\omega_p$ is the plasma frequency (in\units{eV} units) of 
the foil material, and $d_1$ is 
its thickness in\units{\mu m} \cite{Cobb}. At higher $\gamma$ 
the radiation energy increases up 
to a saturation value given by $\gamma_{sat} \sim \gamma_{th} (d_2/d_1)^{1/2}$ 
\cite{Artr}, where $d_2$ is the width of the gap between the 
foils. 

Similar behaviour has 
also been observed for irregular radiators such as carbon 
compound foam layers or fiber 
mats \cite{Prin,Bung}, where the role of the thin foil is 
played by the cell wall and by 
the fiber element respectively, and the gap by the cell 
pore and by the fiber spacing. One 
important advantage of these materials is their low cost. In 
addition, their densities, and 
consequently the cell or fiber sizes and spacings, can be 
easily selected to produce 
increasing transition radiation in the Lorentz factor range 
$10^3 < \gamma < 10^4$, 
corresponding to a 100\units{GeV} to 1\units{TeV} energy region for 
muons. We have tested a variety of 
these materials, trying to obtain the maximum photon yield with 
minimum radiator thickness, 
while maintaining at the same time the widest range between 
$\gamma_{th}$ and $\gamma_{sat}$ \cite{noi2}.

Gaseous chambers working in the proportional region are 
generally preferred to solid state 
or scintillation counters for detection of transition 
radiation. In fact, the radiating 
particle, if not deflected by magnetic fields, releases its 
ionization energy in the same 
region as the X-ray photons, introducing a background signal 
that can be reduced if a 
gaseous detector is used. The gas must provide efficient 
conversion of the TR photons, 
leading to the use of high-Z gases such as argon, krypton, or 
xenon. Multiple module TRDs, 
with optimized gas layer thickness, are normally employed to 
improve the background 
rejection. A reduced chamber gap limits the particle 
ionizing energy losses, while those 
X-rays escaping detection may be converted in the downstream chambers.

The measurement of TR using proportional chambers is 
generally based on one or both of 
two methods:
\begin{itemize}
\item the ``charge measurement'' method, where the signal collected from
a chamber wire is amplified with a time constant of a few hundred ns and 
then charge analyzed by ADCs \cite{Fisc};
\item the ``cluster counting'' method,
where the wire signal is sharply differentiated in order to discriminate  
the $\delta$-ray background from the clusters of ionization from X-ray 
photoelectrons producing pulses (hits) exceeding a threshold amplitude 
\cite{Fabj}.
\end{itemize}
In each case a cut on the analyzed charge or on the number of 
clusters discriminates radiating particles from slower nonradiating ones.

\subsection{Detector description}
We have built three TRD modules, each of about 12\units{m^2} 
surface area for the MACRO experiment \cite{mac1,mac2} at the Gran Sasso 
Laboratory (LNGS). The laboratory is located at an average depth of 
3700\units{hg/cm^2}, with a minimun depth of 3200\units{hg/cm^2}.
The differential distribution of the  residual energy of the 
downgoing muons is expected to 
be nearly flat up to 100\units{GeV}, falling rapidly in the 
\units{TeV} region. The 
mean muon energy is a 
few hundred \units{GeV} \cite{Berg}.
The TRD was designed to explore the muon energy range from 
100\units{GeV} to 1\units{TeV}. Below this 
energy range there is no TR emission for the radiator parameters chosen.
In the range 0.1--1\units{TeV} the response versus $\gamma$ is
approximately linear. For energies greater than 1\units{TeV}, where the 
muon flux is estimated to be a few percent of the total, 
the TR response is saturated.

In order to study the energy spectrum of multimuon events, a large 
area TRD with relatively 
fine spatial resolution is required. The total multiple muon event 
rate for MACRO is 
roughly 0.015\units{Hz}, and the average separation of muons within an
event is of the order of a 
few meters \cite{Mac3}. 
In order to obtain a reasonable sample of these events 
a detector with an area of several tens of square meters is needed.

For the TRD active detector we have adopted 6 meter long 
proportional counters  
having a $6 \times 6$\units{cm^2} square cross section. The polystyrene 
walls of the counters 
are slightly thinner than 1\units{mm}. 
The proportional tube cross section of $6 \times 6$\units{cm^2} is a 
compromise  between 
efficiently converting the TR photons in an argon-based gas mixture, 
while at the same time 
maintaining the ionization energy 
loss of the muon at a relatively low level. The design parameters were 
checked by calculations based on a Monte Carlo \cite{noi1} 
and from tests in a pion/electron beam at energies 1--5\units{GeV}, 
covering the Lorentz factor 
interval $10^3 < \gamma < 10^4$ \cite{spin}.

A layer of these counters is placed between each radiator layer, forming 
a large multiple 
layer TRD. The TRD units were installed on the floor of the
upper MACRO detector with 
the proportional counters running parallel 
to the streamer tubes, 
simplifying the track reconstruction. The number of TRD layers 
was fixed at ten in order 
to constrain the number of channels, and to take into account the
2 meter maximum 
available height for a detector inside MACRO. 
The radiator thickness was limited for the same reason to 10\units{cm}.
Each TRD module has an active volume of $6 \times 1.92 \times 1.7$\units{m^3} 
and contains 32 tubes per layer, interleaved with the foam 
radiators. The bottom tube layer 
is placed on an eleventh radiator. In this way, the detector is 
symmetric with respect to 
downgoing and upgoing muons, thus offering the additional 
opportunity for measuring the 
energy of neutrino induced upgoing muons.

The radiator material used was Ethafoam 220, having a density of 35\units{g/l},
and cells of approximately 0.9\units{mm} diameter and 35\units{\mu m} wall 
thickness \cite{Fabj77}. These cell dimensions provide a relatively wide 
range between $\gamma_{th}$ and $\gamma_{sat}$. 
The TR spectra from 
Ethafoam of equivalent density have already been measured by many authors  
\cite{Prin,Fabj77,Cher2} and 
match properly with the transmission characteristics of the 
proportional tube wall.

A reduced scale prototype exposed to a pion/electron test beam was 
used to determine the 
response function of the detector, and to develop and test the TRD 
readout electronics.  In 
two recent 
papers \cite{noi2,spin} we have analyzed the behavior of the TR 
energy versus  $\gamma$ by 
the method of charge analyzing the signal, and, in addition, 
we have  investigated the 
dependence of the number of TR photons versus  $\gamma$. We 
found that the dependence 
on $\gamma$ of the number of photons is quite similar to that of the TR 
energy, as has been previously reported by other authors \cite{Deni}. 
Therefore, we have equipped the TRD with cluster counting 
electronics, since this method 
has proven to be more reliable and less 
expensive than the ``charge measurement'' method.

The total cluster count (total number of hits) measured in the TRD 
follows a Poisson distribution with an average number of hits 
of the order of ten. In Fig.~\ref{test} 
we show the average number of hits for Ethafoam 
at various $\gamma$ and beam crossing angles.
The average number of hits obtained from electrons without 
radiators is indicated for normal incidence. The response curves
show a behavior compatible with the relativistic rise 
($\gamma < 100$) and the Fermi 
plateau for the energy loss of a fast particle.

In Fig.~\ref{evd} we show a computer display of a multi-muon event 
in the MACRO/TRD 
detector. The muons enter MACRO from the top, pass through the 
TRD, and then exit through 
the lower MACRO detector. 
The TRD readout trigger is provided by the MACRO muon trigger \cite{mac2}.
In this display the number of hits produced by the muons are
indicated by different symbols.

\section{Data selection}

In this analysis we consider the data collected from April 1995 to 
August 1997 by the first 
TRD module. A selection was made to disregard those MACRO runs 
in which the TRD was 
affected by stability problems or was malfunctioning. We started
with a raw data sample of 4665 runs, in which 
215184 muons entered the TRD.  This initial sample consisted of 185915 
single muons, 19875 double muons 
and 9394 muons in events of high multiplicity. Since the TRD calibration 
was performed with particles 
crossing  all ten detector layers and at zenith angles below 
45\degrees~\cite{spin}, in the present 
analysis only single muons meeting these constraints have been 
included. Runs having muon 
rates more than 
three standard deviations with respect to the average have been excluded.

To evaluate the muon energy, we sum the number of TRD hits along 
the straight line fit to 
the track reconstructed by the MACRO streamer tubes
(Fig.~\ref{evd}). The distribution of deviations between the 
reconstructed track and TRD 
hits is Gaussian, with a standard deviation of 
$\sigma$=1.86\units{cm}. In reconstructing a 
track, we consider only the tubes within 3$\sigma$ 
of the track.

In order understand the effects of long term detector gain variations,
we have calculated the average number of hits for single muons 
collected in each run. The 
distribution is Gaussian, with an average number of hits equal to 
4.31 and a standard 
deviation $\sigma$=1.0. 
Those runs with averages fell outside three standard deviations from 
the mean have been excluded. The excluded runs suffered from 
gas gain drifts or from occasional power failures. The final 
data sample consists of 60256 
single muons, for a livetime of about 560.5 days. The reduction 
of this sample to roughly 
1/3 of the raw data sample is mainly due to the requirement 
that the muons cross ten 
TRD planes.

\section{Muon energy spectrum}

In Fig.~\ref{hit} the distribution of the number of hits 
in the single muon tracks in the final event sample is shown. The 
slope change which occurs 
at roughly n$_{hits}$ = 15 is due to the TRD response saturation 
at an energy of about 1\units{TeV}. 
This distribution is then used to obtain the single muon 
energy spectrum.  

We have used an unfolding technique, following the 
prescriptions of refs.\cite{Dag,Maz}.  
Unfolding methods require that the 
distribution must be limited to a finite interval. When this 
condition is not fulfilled, 
as for the cosmic ray energy spectrum, the method cannot be automatically 
applied. However, 
in our case the detector response is flat outside the 0.1--1\units{TeV} 
energy interval, thus 
ensuring that the measured quantity, namely the number of hits, becomes 
effectively ``bounded''.

\subsection{Detector response}

\begin{figure}
\resizebox{15.0cm}{15.0cm}{\includegraphics{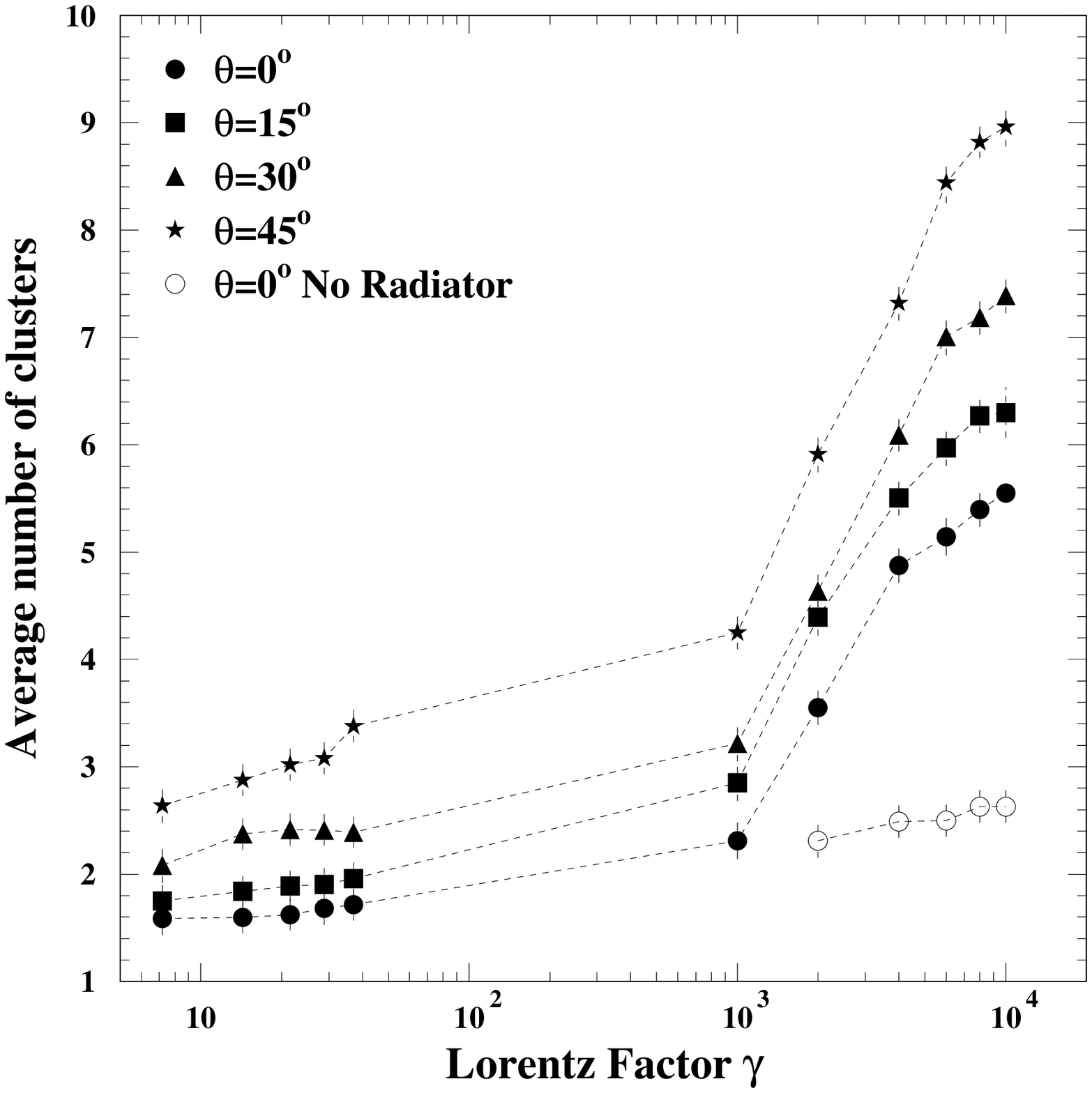}}
\caption{Average number of hits plotted versus the Lorentz
factor $\gamma$ for several beam crossing angles. Dots: 0\degrees{}
incident beam angle; open circles: 0\degrees{} beam angle without radiator; 
squares: 15\degrees{} beam
angle; triangles: 30\degrees{} beam angle; stars: 45\degrees{} beam angle.
The dashed lines are drawn to guide the eye.}
\label{test}
\end{figure}

\begin{figure}
\resizebox{15.0cm}{15.0cm}{\includegraphics{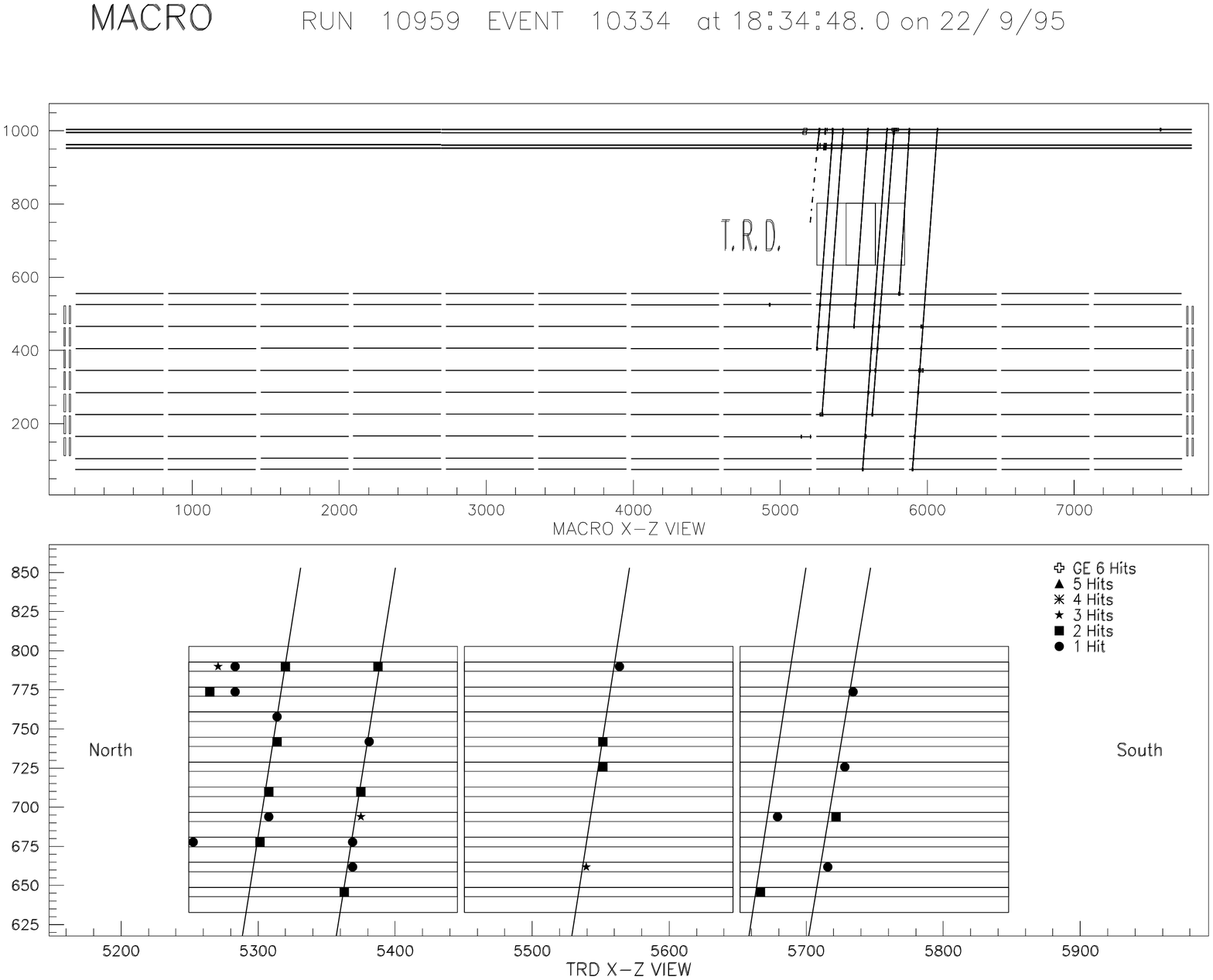}}
\caption{Display of a multiple muon event crossing MACRO and the TRD.
The upper part of figure shows the whole MACRO detector in the 
view orthogonal to the streamer tubes, while in the lower part 
only the TRD in the view orthogonal to proportional tubes is shown.
The number of hits produced in the TRD are shown by
different symbols.
While the second muon from the right has $E_{\mu}$ approximately 
200\units{GeV},
the other muons have energies of roughly 500\units{GeV}.}
\label{evd}
\end{figure}

\begin{figure}
\resizebox{15.0cm}{15.0cm}{\includegraphics{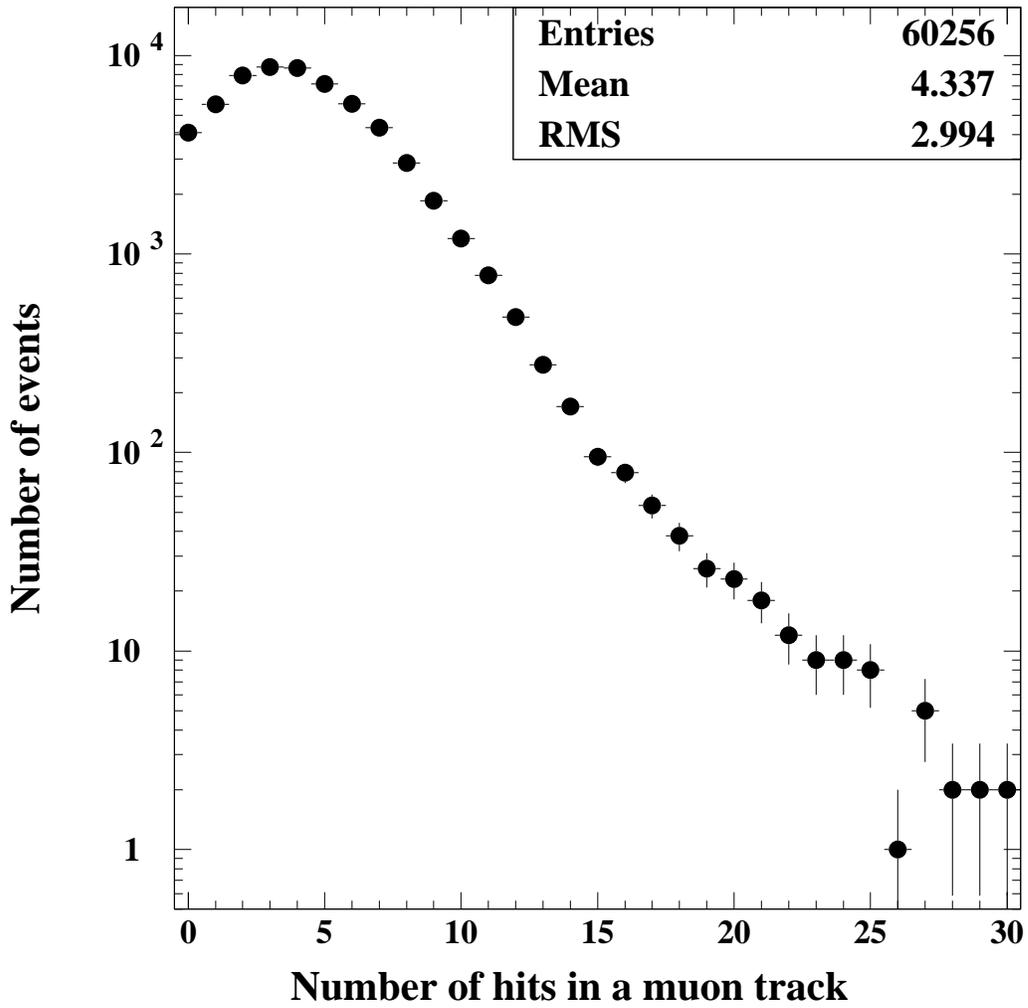}}
\caption{Hit distribution for single muon tracks crossing the 10 TRD
planes with zenith angles less than 45\degrees{}. 
Only statistical errors are shown.}
\label{hit}
\end{figure}

The distribution of the hits collected along a muon track by the TRD at 
a given zenith and azimuth angle, $N(k,\theta,\phi)$, 
can be related to the residual energy 
distribution of muons, $N(E,\theta,\phi)$, by
\begin{equation}
\label{eq:e1}
N(k,\theta,\phi) = \sum_{j} p(k \mid E_j,\theta,\phi) 
N(E_j,\theta,\phi)
\end{equation}
where the detector response function, $p(k \mid E_j,\theta,\phi)$,
is the probability to observe $k$ hits in a track of a given 
energy $E_j$ and at a given angle $\theta$ and $\phi$.
This response function must 
contain both the detector acceptance and the 
event reconstruction efficiency. We derived this function
by simulating MACRO using GEANT 
\cite{Brun}, including the simulation of trigger efficiency.
The TRD simulation was based on the test beam calibration data \cite{spin}
(Fig.~\ref{test}).

As shown in Fig.~\ref{test}, the TRD exhibits a different behavior 
in different energy 
regions. It provides a flat response below 100\units{GeV}, 
a linear increasing response up to about 1\units{TeV}, and then saturates. 
The energy bins used in presenting the muon energy spectrum were chosen 
on the basis of 
this behavior, and on the basis of the momentum bins used 
in the calibration runs. The 
first bin  covers the energy range from 0 to 50\units{GeV}, while 
the last represents a lower 
limit at 1\units{TeV}
corresponding approximately to the TRD saturation energy for 
muons. On the same basis we 
have chosen four angular bins from 0 to 45 degrees.

The detector response function was derived using an unbiased muon 
energy spectrum, i.e., one which was flat versus energy, $\theta$ 
and $\phi$. It was 
calculated by taking the ratio of the number of events 
producing $k$ hits at a given energy 
and incident angle $\theta$ to the total number of the events 
in the same energy bin and 
incident angle.
The simulated data were produced in a form similar to experimental
data, in order to process it with the same analysis procedure.

Low energy muon data was used to verify the consistency of 
the simulation with the 
behavior of the TRD during data taking.
We selected muons with $\gamma < 20$ (corresponding to an average energy
of about 1.5\units{GeV}) which cross the TRD and then stop
in the lower MACRO detector, and muons with large scattering 
angles in the lower part of 
MACRO.  The selection of muons stopping in the MACRO layers
below the TRD was based on considering only tracks crossing less 
than eight out of ten 
layers of the lower MACRO structure.



\begin{figure}
\resizebox{15.0cm}{15.0cm}{\includegraphics{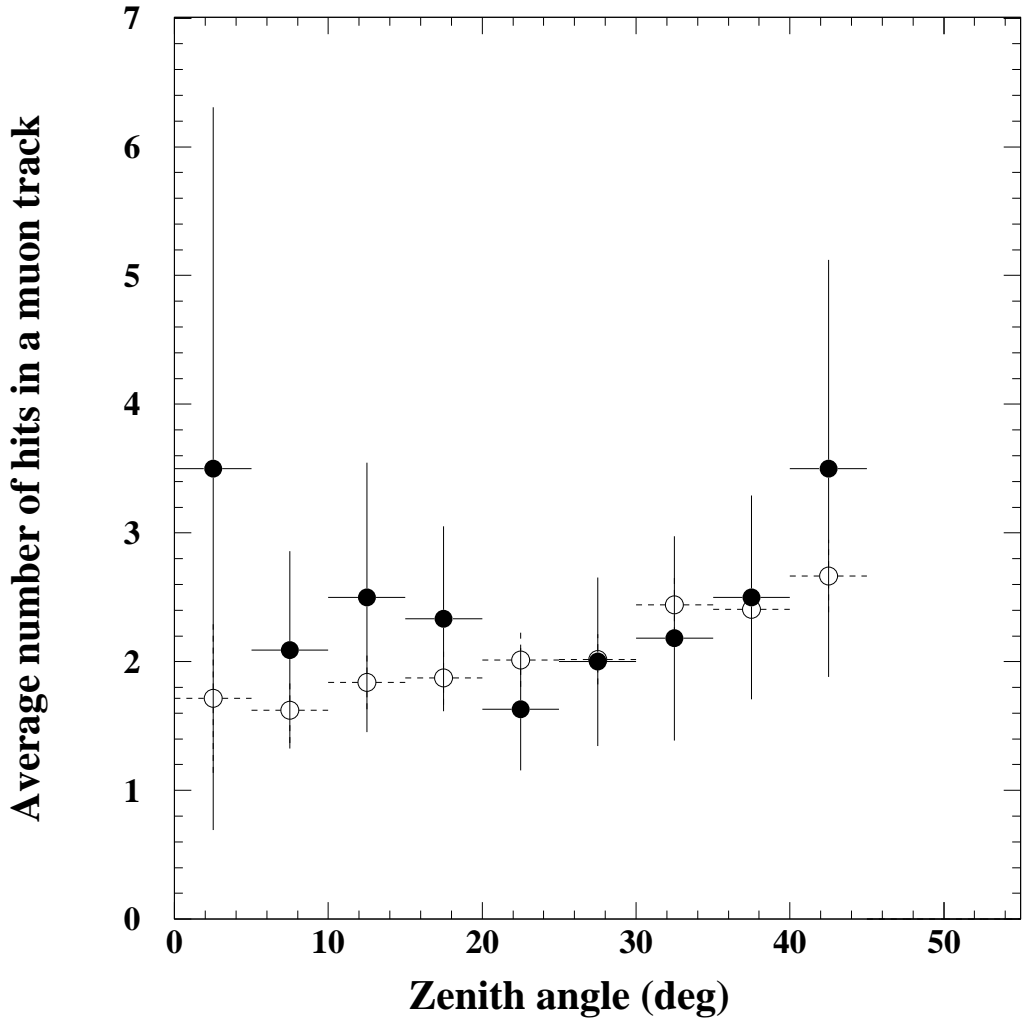}}
\caption{Average number of hits versus zenith angle
for muons crossing the TRD
and stopping in the lower MACRO detector.
Black circles: TRD data; open circles: Monte Carlo simulation.
Only statistical errors are shown.}
\label{fig4b}
\end{figure}

The average number of hits versus zenith angle is shown in 
Fig.~\ref{fig4b} together
with the same average hit distribution simulated by Monte Carlo 
procedure described
above. The experimental
data are in good agreement both with the Monte Carlo and with
the TRD calibration points of the equivalent energy, 
namely for $\gamma < 20$ (Fig.~\ref{test}).


\subsection{Results}

The unfolding procedure described above was applied to the TRD 
experimental data, starting with a trial spectrum
assigned to the unfolded distribution \cite{Dag,Maz} 
according to a local energy spectrum of muons at 4000\units{hg/cm^2}
with a spectral index of 3.7 as reported in \cite{Lipa}:
\begin{equation}
\label{eq:e2}
N_0(E,\theta,\phi) 
\sim e^{- \beta h (\alpha - 1)} (E+\epsilon(1-e^{-\beta h}))^{-\alpha} .
\end{equation}
The parameters are: $h=4$\units{km\,w. e.}, $\alpha=3.7$, 
$\beta=0.383$\units{(km\,w. e.)^{-1}} and $\epsilon=0.618$\units{TeV}.

The iterative procedure of the unfolding method is terminated when the   
reconstructed distribution at the  $i$th iteration is equivalent 
to the previous one at a 
probabilty  $\geq 99\%$. The $\chi^2$ is calculated by summing 
over the squared differences 
between the channel content of two subsequent distributions, normalized 
to the square of 
the statistical errors. The final result is found to be 
unaffected by the choice of the
spectral index in the initial probability function.

\begin{figure}
\resizebox{15.0cm}{15.0cm}{\includegraphics{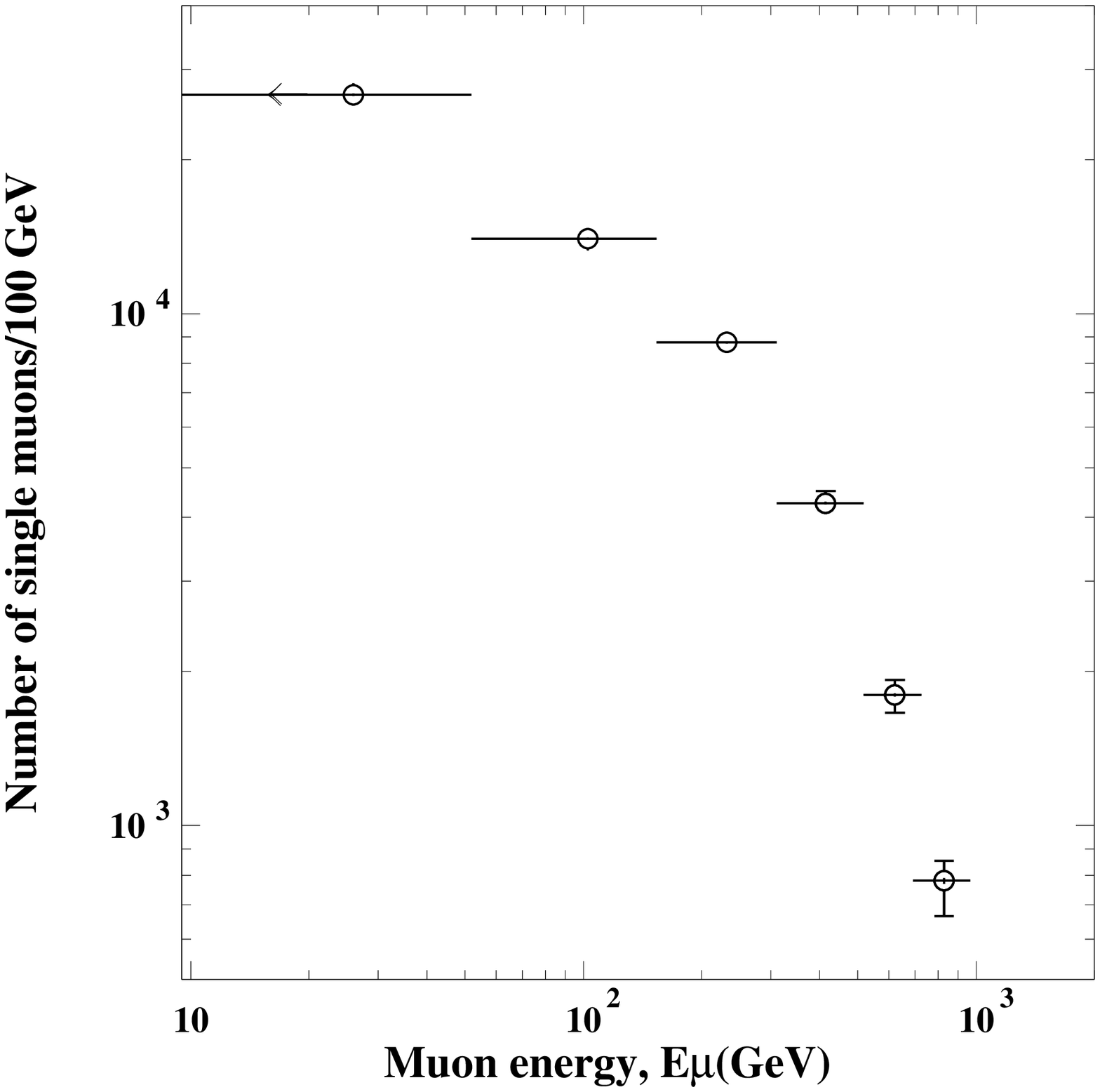}}
\caption{Differential energy distribution of single muons with zenith
angle $\leq$ 45\degrees{} measured with the TRD. The spectrum was obtained
by unfolding the hit distribution shown in Fig.~\ref{hit}.}
\label{diff}
\end{figure}

\begin{figure}
\resizebox{15.0cm}{15.0cm}{\includegraphics{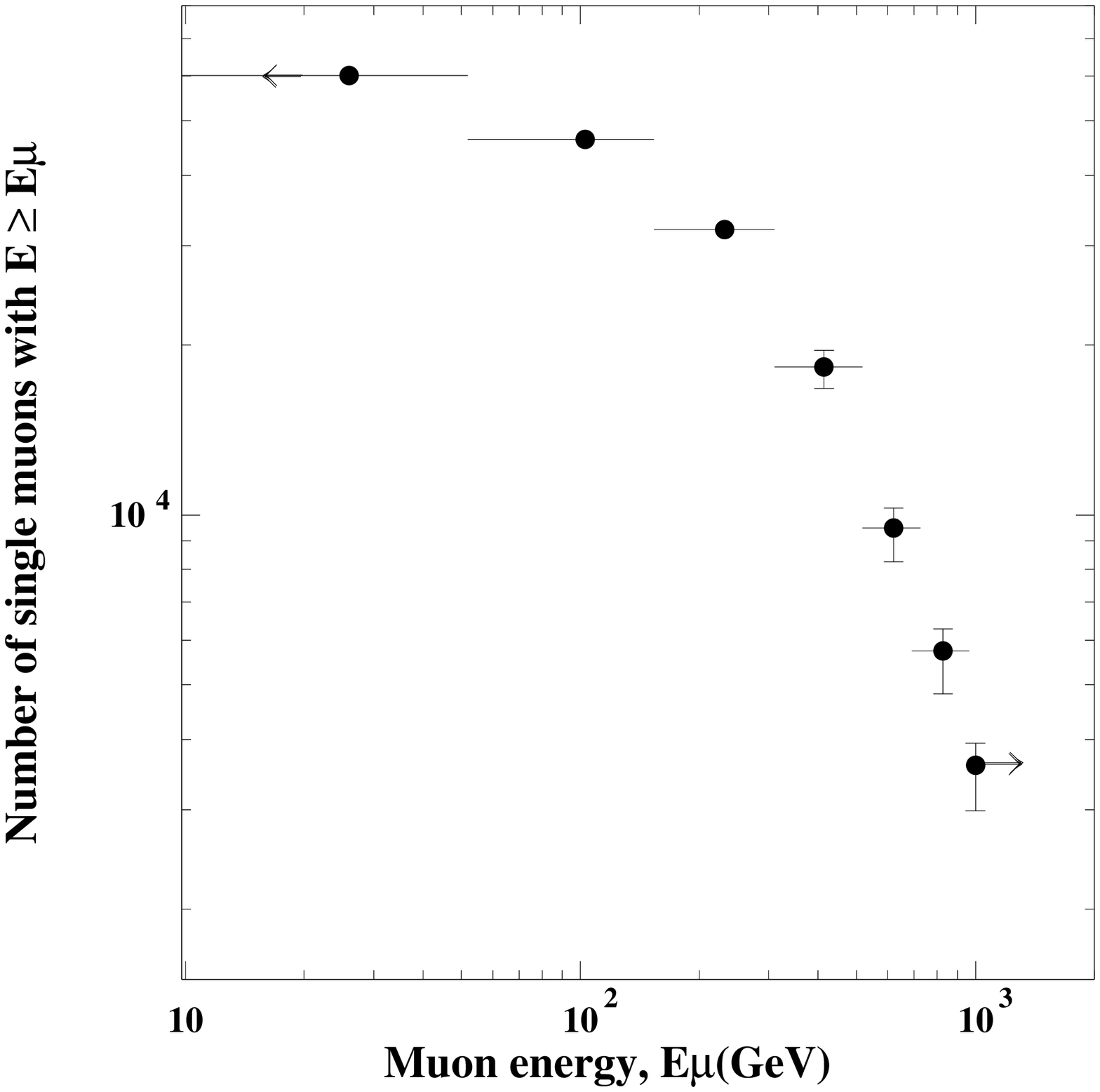}}
\caption{Integral energy distribution of single muons with zenith
angle $\leq$ 45\degrees{} measured with the TRD. The spectrum was obtained
by unfolding the hit distribution shown in Fig.~\ref{hit}.}
\label{inte}
\end{figure}

\begin{figure}
\resizebox{15.0cm}{15.0cm}{\includegraphics{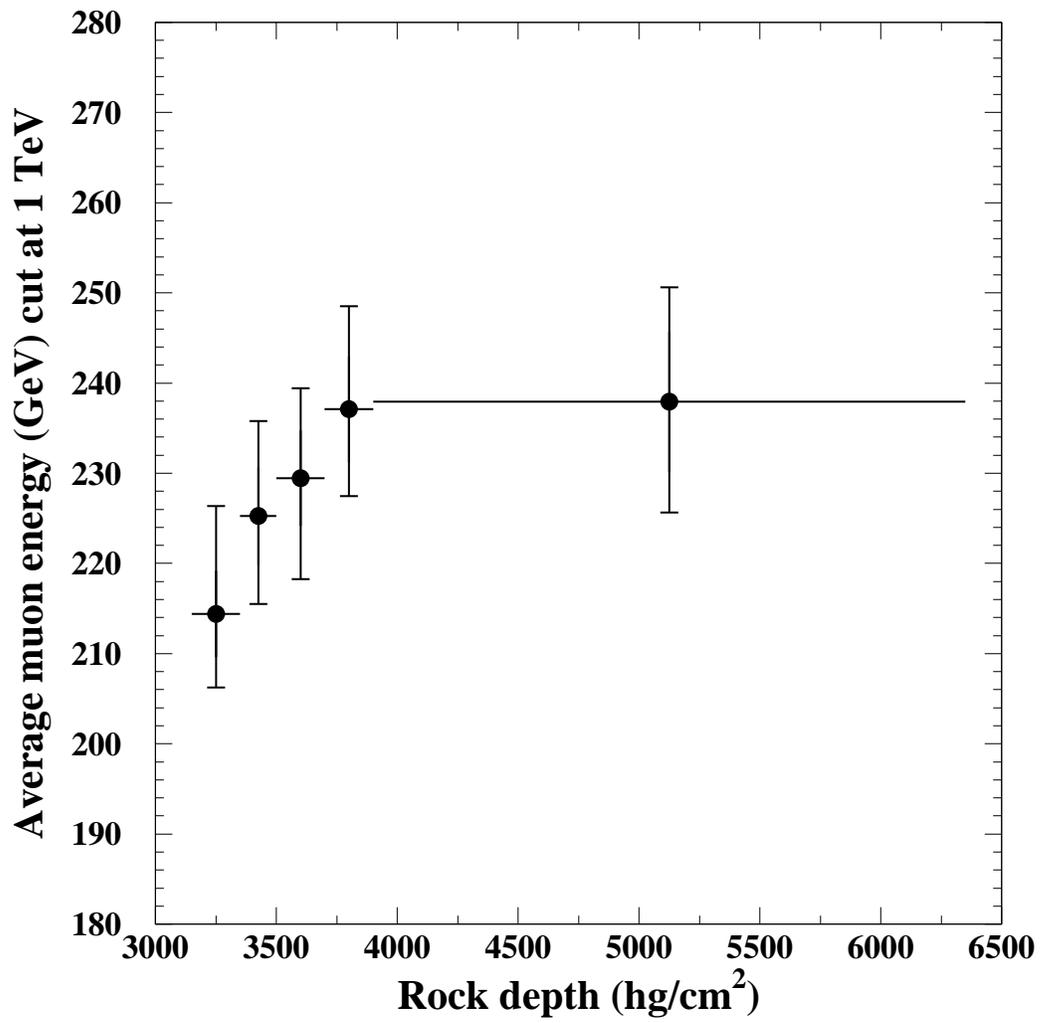}}
\caption{Average single muon energy, computed with a cut at 1\units{TeV},
versus the standard rock depth.}
\label{ecut}
\end{figure}

\begin{figure}
\resizebox{15.0cm}{15.0cm}{\includegraphics{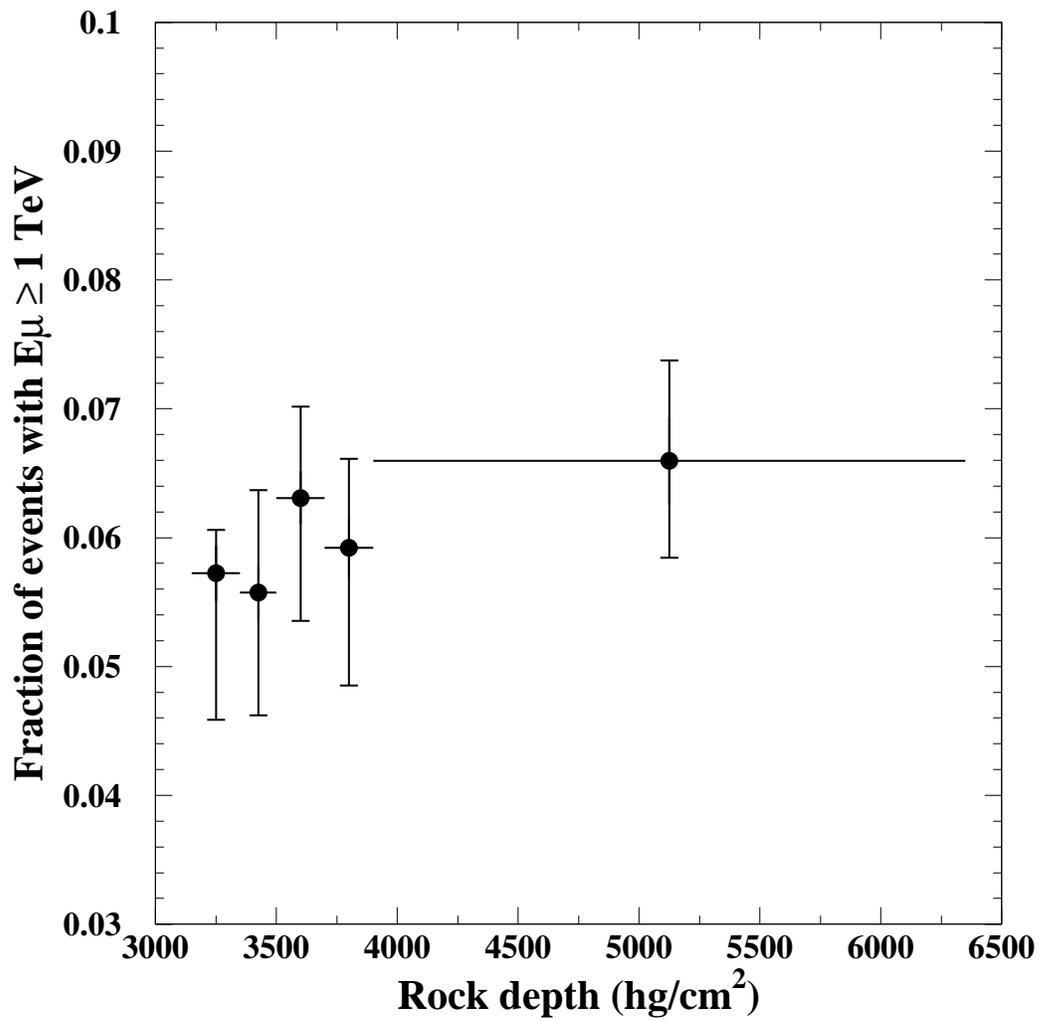}}
\caption{Fraction of single muons with energy greater than 1\units{TeV}
versus the standard rock depth. }
\label{frac}
\end{figure}

In Figs.~\ref{diff} and~\ref{inte} the muon energy differential spectrum and
the muon energy integral spectrum are reported.  Fig.~\ref{ecut}
shows the average energy of events below 1\units{TeV} versus rock 
depth, while Fig.~\ref{frac} 
shows the fraction of muons with energies exceeding 1\units{TeV} 
versus rock depth. The fraction 
is about 6$\%$, independent of rock depth. 
A topographic map of the terrain above MACRO was used to obtain the
rock depth from the direction of the muon track.
The average muon energy  in 
the energy range $0.1 < E < 1$\units{TeV} is 225 $\pm$ 3\units{(stat.)} 
$\pm$ 4\units{(syst.)} \units{GeV}. 
The quoted systematic errors are due to beam calibration
uncertainties, estimated at $\pm$ 
2\%. They have been obtained by changing 
the calibration input data in the unfolding procedure by the 
same percentage. 
The statistical and systematic errors  have 
been added in quadrature in the figures.

The single muon spectrum deep underground is determined by the spectrum
at the surface and by the energy losses in the rock.
In this analysis we have investigated the consistency of the
residual muon energy spectrum with the
``all-nucleon'' energy spectrum of primary cosmic rays.
We have compared our measurements to the predictions from two extreme
hypotheses on the primary spectra \cite{Cmac} assuming a given range
of the spectral index, 
namely the ``Light'' (i.e., proton-rich) \cite{Ligh} and the 
``Heavy'' (i.e., Fe-rich) \cite{Heav} compositions. 
In the present analysis we have adopted a normalization procedure for these
compositions in order to reproduce the known abundances and spectra
directly measured, and to match the extensive air shower data
at higher energies \cite{Mac3}. The interaction of 
the cosmic rays in the atmosphere was simulated with the HEMAS 
code \cite{Fort}. 
The secondary muons at sea level were propagated through the rock, with
the muon energy loss in the rock evaluated according 
to the prescriptions of ref.~\cite{Lipa}. 
The rock thickness was calculated at each $\theta$ and 
$\phi$ from the Gran Sasso map 
\cite{Gsro}.
We used the correction procedure described in ref.~\cite{Wrig} for the 
conversion to standard rock.  We find that
our measurements of the average single muon energy and the
fraction of single muons with energy $\geq$ 1\units{TeV} are
in agreement with spectra obtained from the Monte Carlo models.

The experimental average muon energy over all energies was 
calculated by adding to the 
average energy obtained with an energy cut at 1\units{TeV} the 
contribution from muons of greater energy. The high energy 
contribution was estimated by 
multiplying the measured fraction
of muons with energy $\geq$ 1\units{TeV} by the average muon energy 
above 1\units{TeV}:

\begin{equation}
\label{eq:e3}
{<}E_{\mu}{>} = (1-f) \cdot {<}E_{\mu}{>}_{cut} + 
		f \cdot {<}E_{\mu}{>}_{no cut}
\end{equation}
where $f$ is the fraction of events with $E \geq 1$\units{TeV} (measured), 
${<}E{>}_{cut}$ is the average energy with $E < 1$\units{TeV} (measured) and
${<}E{>}_{no cut}$ is the average energy with $E \geq 1$\units{TeV}.

The evaluation of ${<}E{>}_{no cut}$ was based on a simple extrapolation 
of the 
local energy spectrum as reported in Eqn.~(\ref{eq:e2}) using the
same parameters $\alpha=3.7$, $\beta=0.383$\units{(km\,w. e.)^{-1}} 
and $\epsilon=0.618$\units{TeV} for the
depth interval shown in Figs.~\ref{ecut} and~\ref{frac}.
The average muon 
energy obtained in this way is $320 \pm 4(stat.) \pm 11(syst.)$ \units{GeV} 
and does not
change appreciably with variation of these parameters.
A variation of 3\% in the above parameters, as is typically quoted by
various authors (e.g., ref.~\cite{batr}), implies uncertainties of about
0.1\% for $\beta$, 0.2\% for $\epsilon$ and 1\% for $\alpha$.
These uncertainties are significantly less than
our quoted error.

\begin{figure}
\resizebox{15.0cm}{15.0cm}{\includegraphics{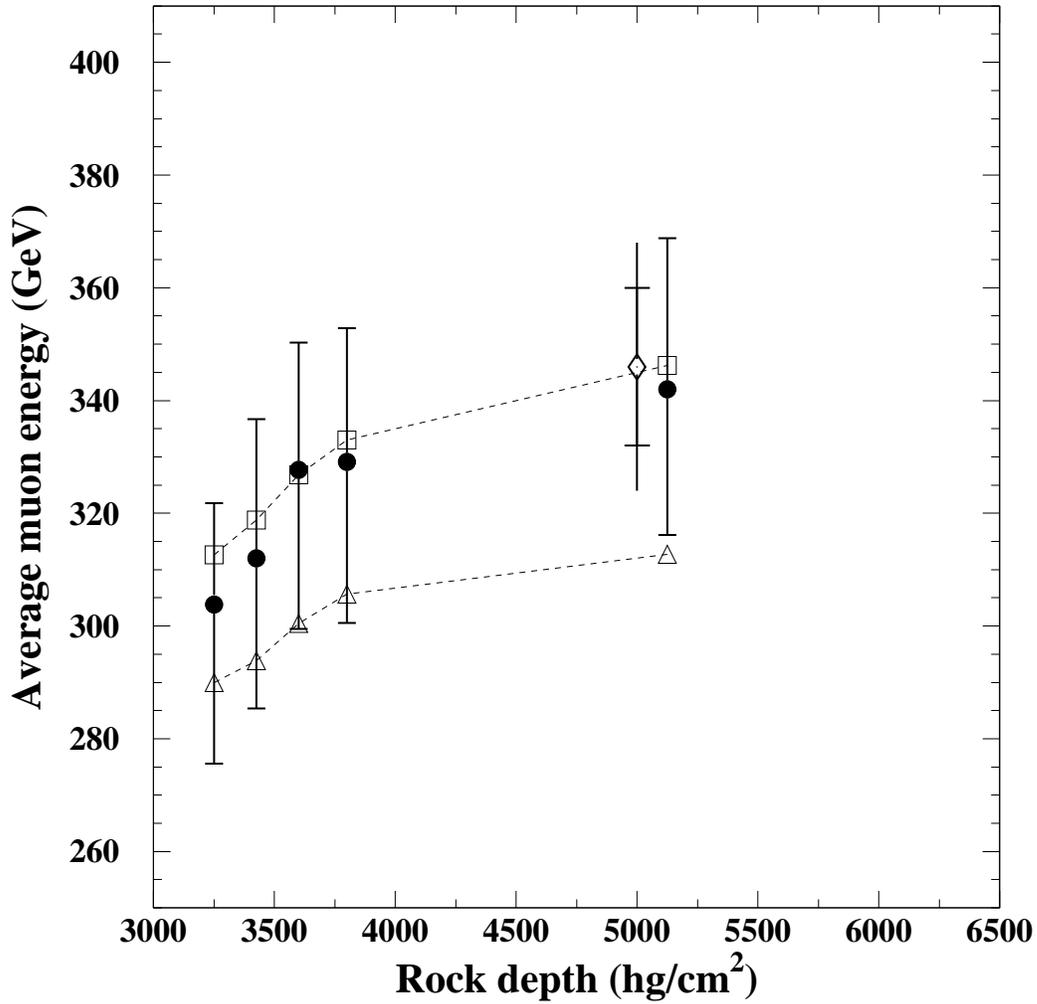}}
\caption{Average single muon energy measured by the MACRO TRD 
(black circles) versus standard rock depth. 
The open symbols connected by dashed lines are the predictions of a
HEMAS-based Monte Carlo for the ``Light'' (squares) and 
``Heavy'' (triangles) composition models.
The result reported by the NUSEX experiment is shown by the diamond
(the extensions of the error bar represents the 
systematic uncertainty added in quadrature to the statistical error
[4]).}
\label{etot}
\end{figure}

Fig.~\ref{etot} shows the average single muon energy as a function of
rock depth. Also shown are the predictions of the two 
composition models studied. The 
NUSEX experimental point is also shown, and is in good 
agreement with our measurements.  
The present result is not able to discriminate between 
the two composition models.   

\section{Conclusions}
We have measured directly the residual energy of cosmic ray muons 
at the Gran Sasso underground laboratory, using 
a TRD which has been operational since April 1994. 
The average single muon energy, in the range 0.1--1\units{TeV}, is
225 $\pm$ 3 (stat.) $\pm$ 4 (syst.)\units{GeV}. The fraction of muons with 
energies $>$ 1\units{TeV} 
is 6.0 $\pm$ 0.1 (stat.) $\pm$ 0.4 (syst.)\% in the depth 
range 3150--6500\units{hg/cm^2}. Treating the events with energies 
greater than 1\units{TeV} in the 
manner described above, the average
single muon energy in this depth range is 
320 $\pm$ 4 (stat.) $\pm$ 11 (syst.)\units{GeV}.
The results are in agreement 
with the calculations of the energy loss of the muons in 
the rock above the detector.

\newpage

\newpage

\newpage


\newpage

\newpage

\newpage

\newpage

\newpage

\newpage

\end{document}